\documentclass[12pt]{article}
\usepackage{epsfig,a4}
\usepackage{amssymb}
\usepackage{amsmath}
\usepackage{indentfirst}
\usepackage{graphicx}
\usepackage{graphics}
\usepackage{epsfig}
\usepackage{multirow}
\usepackage{color}
\textheight 23.0cm \textwidth 16.5cm \def\ol{\overline}
\def\mP{{\Omega'< m}}
\def\mm{{m\leq\Omega'}}
\def\wt{\widetilde}
\begin{document}
\title{ Color symmetric superconductivity in a
phenomenological QCD  model }
\author{
Henrik Bohr\\
{\it \small Department of Physics, B.307, Danish Technical
University,}\\{ \it \small DK-2800 Lyngby, Denmark}\\Constan\c{c}a
Provid\^encia,
Jo\~ao da Provid\^encia\\
{\it \small CFC, Departamento de F\'\i sica, Universidade de Coimbra,}\\
{\it \small  P-3004-516 Coimbra, Portugal}  }


\maketitle

\def\e{\rm e}\def\d{{\rm d}}
\def\oln{\overline{n}}
\def\ve{\varepsilon}
\def\bfp{{\bf p}}
\def\bfx{{\bf x}}
\def\bfk{{\bf k}}
\def\bfq{{\bf q}}
\def\wtn{\widetilde{n}}
\def\wtt{\widetilde\theta}
\abstract{ In this paper, we construct a theory of the NJL-type
where superconductivity is present, and yet the super-conducting
state remains, in the average, color symmetric. This shows that the
present approach to color superconductivity is consistent with color
singlet-ness. Indeed, quarks are free in the deconfined phase, but
the deconfined phase itself is believed to be a color singlet.  The
usual description of the color superconducting state violates color
singlet-ness. On the other hand, the color superconducting state
here proposed, is color symmetric in the sense that an arbitrary
color rotation leads to an equivalent state, with precisely the same
physical properties.}
\section{Introduction}

It is presently accepted that  quark and gluon fields are the
building blocks of hadronic matter, in the framework of quantum
chromodynamics (QCD). The investigation of the phase structure of
hadronic matter is a topic of great current interest. A diversity of
phases is expected at high densities: chiral-symmetry restoration,
deconfinement and color-superconductivity. Since, due to the
complexity of the theory, it is extremely difficult, if not
impossible, to obtain exact results directly from QCD when
perturbation theory cannot be applied, effective models, such as the
Nambu-Jona-Lasinio (NJL) model \cite{NJL,klevansky,hatsuda}, must be
employed and have been used with great success to investigate the
properties of hadronic matter and to develop insight into its phase
diagram \cite{buballa}.

Recently, the color superconducting phase in quark matter has been
investigated by many authors \cite{alford}, in the framework of the
Bardeen-Cooper-Schrieffer (BCS) approach, which is familiar from
condensed matter physics. For a recent review, see \cite{alford1}.
Quarks are free in the deconfined phase, but the deconfined phase
itself is expected to be a color singlet. In Refs.
\cite{gerhold,dietrich} it has been argued that in QCD the
superconducting phase is automatically color symmetrical.  Our basic
assumption is, therefore, the existence of globally color-symmetric
superconducting phases, and our aim is to discuss how these phases
may be described in terms of effective models with 4 fermion
interactions, such as the NJL-like model considered in Refs.
\cite{alford,alford1}, for which it is assumed that the gluon
degrees of freedom have been integrated over.


A BCS state $|\Phi\rangle$ describes a physical state with zero net
color charge if $N_1=N_2=N_3$, where $N_i$ denotes the average
number of quarks of color $i$. This means that
\begin{eqnarray}\langle\Phi|S_{\lambda_3}|\Phi\rangle=\langle\Phi|S_{\lambda_8}|\Phi\rangle=0.\label{cneutral}
\end{eqnarray} Here, $S_{\lambda_k}$ denotes the $SU(3)$ generator associated
with the Gell Mann matrix ${\lambda_k}$. However, the requirement
(\ref{cneutral}), which is implemented in \cite{iida,buballa1}, is
not sufficient to insure that $|\Phi\rangle$ is physically
acceptable. A stronger condition must then be imposed. Indeed, the
$SU(3)$ symmetry, being a gauge symmetry, cannot be broken,
according to the discussion in \cite{gerhold,dietrich}, so that
color rotated BCS states must be equivalent in the sense of the
physics they describe. Let $U_c$ denote an arbitrary color rotation,
i.e., $U_c=\exp \sum_{k=1}^8ix_kS_{\lambda_k}$, the parameters $x_k$
being arbitrary and real. The BCS state $|\Phi\rangle$ must be
equivalent to the state $U_c|\Phi\rangle$, for any $U_c$, as far as
expectation values of physical observables are concerned. Therefore,
the condition (\ref{cneutral}) must be replaced by
\begin{eqnarray*}\langle\Phi|U_c^\dagger
S_{\lambda_3}U_c|\Phi\rangle=\langle\Phi|U_c^\dagger
S_{\lambda_8}U_c|\Phi\rangle=0,\end{eqnarray*} for an arbitrary
$U_c$, and this implies
\begin{eqnarray}\langle\Phi|S_{\lambda_k}|\Phi\rangle=0,{\rm~~~ for~~~}
k=1,2,\cdots,8.\label{csinglet}\end{eqnarray} This is the condition
the BCS state $|\Phi\rangle$ must satisfy in order to be physically
meaningful. If only the condition (\ref{cneutral}) is implemented,
and not the condition (\ref{csinglet}), the BCS state $|\Phi\rangle$
is, in general, not equivalent to the state $U_c|\Phi\rangle$, so
that it describes a state belonging to a representation of $SU(3)$
other than the singlet one, which is physically unacceptable. In
\cite{bohr} it is shown how the condition (\ref{csinglet}) may be
easily implemented.

In the present paper, we apply to the NJL model the new BCS approach
developed in Ref. \cite{bohr} which uses the generalized Bogoliubov
transformation and leads to a color symmetric BCS vacuum. In Section
2, a mean-field constrained Hamiltonian appropriate for the
description of color superconductivity is presented. In section 3 we
compare the new superconducting state which satisfies
(\ref{csinglet}), with  the usual one, which is not required to
satisfy (\ref{csinglet}). In Section 4, we draw some conclusions.
\section{Color symmetrical superconductivity}

We wish to focus on the diquark condensate
$\langle\bar\psi^Ci\gamma_5\lambda_j\tau_2\psi\rangle,\,j\in\{2,5,7\},$
and, for simplicity, we will neglect the quark-antiquark condensate
$\langle\bar\psi\psi\rangle$.  The notation is the usual one, the
superscript $C$ denoting charge-conjugation. In this sense, the
model is not fully realistic, but is adequate for the present
development. We assume that the pairing interaction is antisymmetric
in the color indices and in the iso-spin indices, i.e. it involves
the Gell-Mann matrices $\lambda_j,\,j\in\{2,5,7\}$ and the iso-spin
matrix $\tau_2$  associated with flavor, when 2 is the number of
flavors. Having in mind a Hamiltonian of the NJL \cite{klevansky}
type, the mean-field constrained Hamiltonian reads:

\begin{eqnarray}\hat K_{MFA}&=&\int \d^3\bfx\left[\bar\psi(\vec
p\cdot\vec\gamma+
M-\mu\gamma_0)\psi+{1\over2}\sum_{j\in\{2,5,7\}}(\Delta^*_j\,\bar\psi^C
i\gamma_5\tau_2\lambda_j\psi+h.c.)\right.\nonumber\\&+&\left.
\sum_{j\in\{2,5,7\}}{|\Delta_j|^2\over4G_C}\,\right],
\end{eqnarray}
where \begin{equation}\Delta_j=-2G_C\langle\bar\psi^C
i\gamma_5\tau_2\lambda_j\psi\rangle, \end{equation}  denotes the BCS
gap. We use the symbol $\hat K$, instead of the more usual symbol
$\hat H$ to stress that this Hamiltonian is constrained in the sense
that it fixes the Fermion number through the chemical potential
$\mu$, which behaves as a Lagrange multiplier. Thus the expectation
value of $\hat K_{MFA}$ is the 
thermodynamical potential which is equal to $-PV$, where $P$ is the
pressure and $V$ the volume, and determines the equation of state.
By $\langle X\rangle $ we denote the average of $X$ in the BCS
vacuum which will be specified in the following. The notation is
essentially the same as in \cite{koide}, slightly modified, as is
required in order to treat the 3 colors on the same footing. In
momentum space we have,
\begin{eqnarray}\hat K_{MFA}&=&\sum_{\bfp,\eta,j,\tau}\left[
(\sqrt{p^2+M^2}-\mu)c^\dagger_{\bfp,\eta,j,\tau}c_{\bfp,\eta,j,\tau}
+(\sqrt{p^2+M^2}+\mu)\tilde c^\dagger_{\bfp,\eta,j,\tau}\tilde
c_{\bfp,\eta,j,\tau}\right]\nonumber\\&
+&{1\over2}\sum_l\Delta_l\sum_{\bfp,\eta,j,k,\tau,\tau'}
(c^\dagger_{\bfp,\eta,j,\tau}c^\dagger_{-\bfp,\eta,k,\tau'}+\tilde
c_{\bfp,\eta,j,\tau}\tilde c_{-\bfp,\eta,k,\tau'})
\epsilon_{jkl}\epsilon_{\tau\tau'}\zeta_{\bfp,\eta}+h.c.\nonumber\\&+&V\sum_l{|\Delta_l|^2\over4G_C},
\end{eqnarray}
where $c^\dagger_{\bfp,\eta,j,\tau}$ and $\tilde
c^\dagger_{\bfp,\eta,j,\tau} $ create, respectively, a quark and an
antiquark of momentum $\bfp$, helicity $\eta$, color index $j$ and
isospin index $\tau$, $
\zeta_{\bfp,\eta}=-\zeta_{-\bfp,\eta}=-\zeta_{\bfp,-\eta}=\zeta_{-\bfp,-\eta},
~|\zeta_{\bfp,\eta}|=1$, and
 \begin{equation}\Delta_l^*=- V^{-1}\,2 G_C\sum_{\bfp,\eta,j,k,\tau,\tau'}
\left[\left(\langle
c^\dagger_{\bfp,\eta,j,\tau}c^\dagger_{-\bfp,\eta,k,\tau'}\rangle
+\langle\tilde c_{\bfp,\eta,j,\tau}\tilde
c_{-\bfp,\eta,k,\tau'}\rangle\right)
\epsilon_{jkl}\epsilon_{\tau\tau'}\zeta_{\bfp,\eta}\right].
\end{equation}
To be precise, the index $j$ in $c^\dagger_{\bfp,\eta,j,\tau}$
labels states of the $3$ representation of $su(3)$, while in $\tilde
c^\dagger_{\bfp,\eta,j,\tau}$ it labels states of the $\bar 3$
representation.

For convenience, we  introduce the notation,
$m=(\bfp,\eta,\tau),\;\ol m=(-\bfp,\eta,-\tau)$. Clearly, $\ol{\ol
m}=m$. The BCS vacuum $|\Phi\rangle$ is defined as the state which
is annihilated by the operators $d_{jm}$ such that
\begin{eqnarray}&&d_{1m}=c_{1m}-K_m(c^\dagger_{2\ol m}-c^\dagger_{3\ol m}),
\quad \mP,\nonumber\\&&  d_{1m}=c^\dagger_{1m}+\tilde K_m(c_{2\ol
m}-c_{3\ol m}), \quad \mm, \label{4}
\end{eqnarray} that is, $d_{jm}|\Phi\rangle=0 $.
 The parameters
$K_m,~\tilde K_m$ are real.  The
notation $
\mP$ means symbolically that $m$ is such that
$\sqrt{p^2+M^2}-\mu>0,$ while  $
\mm$ means that $\sqrt{p^2+M^2}-\mu\leq 0$. The expressions for
$d_{2m},~d_{3m}, $ are obtained by circular permutation of the
indices 1,2,3. The transformation (\ref{4}) is not canonical, since
$\{d_{im},d_{jn}^\dagger\}\neq\delta_{ij}\delta_{mn}$, but the
corresponding canonical transformation, which is not needed for the
present purpose, may be easily obtained.

 At this point, a short explanation may be in order, concerning the
generalized Bogoliubov transformation, symmetric in the color
indices, which has been proposed in eq. (25) of the first paper
cited in \cite{bohr}, and which was supposed to ``diagonalize" the
pairing hamiltonian of the Bonn model. However, there is an error in
that equation and the BCS vacuum associated with the Bogoliubov
transformation it describes fails to produce a non-vanishing gap
$\Delta$. That transformation is appropriate to diagonalize a
pairing Hamiltonian which, although seemingly similar, is actually
essentially different from the pairing Hamiltonian of the Bonn
model, since that Hamiltonian is not invariant under $SU(3)$. That
is the reason for the {\it corrigendum} in \cite{bohr}.

The color symmetrical BCS state reads
\begin{eqnarray}|\Phi\rangle=\exp\sum_{j=1}^3\left(\sum_{ \mP}K_m
A^\dagger_{jm} + \sum_{ \mm}\tilde K_m A_{jm}\right)|
0_{\Omega'}\rangle,
\end{eqnarray} where
$$ | 0_{\Omega'}\rangle=\left(\prod_{j=1}^3\prod_{
\mm}c^\dagger_{jm}c^\dagger_{j\ol m}\right)|0\rangle,$$ and
$$ A^\dagger_{1m}=c^\dagger_{2m} c^\dagger_{3\ol m}+c^\dagger_{2\ol m}
c^\dagger_{3 m},
$$ $|0\rangle$ denoting the quark vacuum.
The expressions for $A^\dagger_{2m},A^\dagger_{3m},$ are obtained by
circular permutation of the indices $1,2,3$. For simplicity, pairing
operators involving anti-quarks $\tilde c$ (negative energy states)
are not shown,  but, in principle, their contribution should be
included. In conformity, we will not show the contribution of
anti-quarks to color superconductivity, but it is straightforward to
include that contribution.

We use the notation $\langle
W\rangle=\langle\Phi|W|\Phi\rangle/\langle\Phi|\Phi\rangle$. We
easily find
\begin{eqnarray}\langle c_{im}^\dagger c_{jm}\rangle=
-{K^2_m\over1+3 K_m^2},~i\neq j,\quad \langle c_{jm}^\dagger
c_{jm}\rangle= {2K_m^2\over1+3K_m^2},\quad  
\mP.\label{9}\end{eqnarray} 
On the other hand,
\begin{eqnarray}\langle c_{im}^\dagger c_{jm}\rangle
={\tilde K^2_m\over1+3 \tilde K^2_m},~i\neq j,\quad \langle
c_{jm}^\dagger c_{jm}\rangle= 1-{2\tilde K^2_m\over1+3\tilde
K^2_m},\quad 
\mm.\label{10} \end{eqnarray} The derivation of of eqs. (\ref{9})
and (\ref{10}) is summarized in the Appendix.


The $U(3)$ generators read \begin{eqnarray}S_{ij}=\sum_m
c^\dagger_{im}c_{jm}.\label{U3} \end{eqnarray} Clearly, the
generators $S_{\lambda_k}$ of $SU(3)$ considered in the Introduction
are related to the generators $S_{kl}$. For instance,
$S_{\lambda_1}=S_{12}+S_{21},~S_{\lambda_2}=-i(S_{12}-S_{21}),~S_{\lambda_3}=S_{11}-S_{22},
$ etc.

Since
 \begin{equation}\langle S_{ij}\rangle=-2\sum_{m<\Omega'}{K_m^2\over1+3K_m^2}+2\sum_{m\geq\Omega'}
{\tilde K_m^2\over1+3\tilde K_m^2},\quad i\neq j,\label{Sij}
\end{equation} we may obviously insure that $\langle
S_{ij}\rangle=0,~i\neq j$, which is the condition for color
neutrality, by conveniently choosing $K_m,\,\tilde K_m$.

Next we compute the contractions $ \langle c_{2 m}c_{1\ol
m}\rangle=\langle c_{3 m}c_{2\ol m}\rangle=\langle c_{1 m}c_{3\ol
m}\rangle=\langle c_{2\ol m}c_{1 m}\rangle=\langle c_{3\ol m}c_{2
m}\rangle=\langle c_{1\ol m}c_{3 m}\rangle=:D_m$, where $D_m$ is
real.
We have,
\begin{eqnarray}D_m={K_m\over1+3K_m^2},\quad
\mP;\quad D_m= {\tilde K_m\over1+3\tilde K_m^2},\quad \mm,\label{14}
\end{eqnarray} the derivation of these relations being left to the
Appendix.

We are now able to compute the expectation value of $\hat K_{MFA}$.
We obtain
\begin{eqnarray}&\qquad\langle\hat
K_{MFA}\rangle&=3\sum_{
\mP}\left(\varepsilon_m{2K_m^2\over1+3K_m^2}+2\Delta~{K_m\over1+3K_m^2}\right)\nonumber\\&&+
3\sum_{ \mm}\left(\varepsilon_m\left(1-{2\tilde K_m^2\over1+3\tilde
K_m^2}\right)+2\Delta~{\tilde K_m\over1+3\tilde
K_m^2}\right)+V\,{3\Delta^2\over4G_C},\nonumber\\&\\&\Delta_1=\Delta_2=\Delta_3&=-{2G_C\over
V}\left(\sum_{ \mP}{K_m\over1+3K_m^2}+\sum_{ \mm}{\tilde
K_m\over1+3\tilde K_m^2}\right)=:\Delta,\end{eqnarray}
where $\varepsilon_m$ stands for $\sqrt{p^2+M^2}-\mu$.  It is
convenient to define the angles $\theta_m,~\tilde\theta_m$ such that
$\sin\theta_m=\sqrt{3}K_m/\sqrt{1+3K_m^2},$
$\cos\theta_m=1/\sqrt{1+3K_m^2},$ $\sin\tilde
\theta_m=\sqrt{3}\tilde K_m/\sqrt{1+3\tilde K_m^2},$
$\cos\tilde\theta_m=1/\sqrt{1+3\tilde K_m^2}.$ Then, we have
 \begin{eqnarray}&\langle\hat
K_{MFA}\rangle&=\sum_{ \mP}\left(2\varepsilon_m
\sin^2\theta_m+2\sqrt{3}\Delta\sin\theta_m\cos\theta_m\right)\nonumber\\&&+
\sum_{ \mm}\left(\varepsilon_m\left(3-2\sin^2\tilde\theta_m\right)
+2\sqrt{3}\Delta\sin\tilde\theta_m\cos\tilde\theta_m\right)+V\,{3\Delta^2\over4G_C},\label{17}\\&\qquad
\sqrt{3}\Delta&=-{2G_C\over V}\left(\sum_{
\mP}\sin\theta_m\cos\theta_m+\sum_{ \mm}
\sin\tilde\theta_m\cos\tilde\theta_m\right).\label{18}\end{eqnarray}
Having in mind eq. (\ref{Sij}), the color neutrality constraint
$\langle S_{ij}\rangle=0$ reduces to
\begin{eqnarray}-\sum_{\mP}\sin^2\theta_m+\sum_{\mm} \sin^2\tilde\theta_m=0.\label{cnc}
\end{eqnarray} The extremum condition reads (see Appendix)
\begin{eqnarray}&&\cos2\theta_m={\varepsilon_m-\lambda\over\sqrt{(\varepsilon_m-\lambda)^2+3\Delta^2}},\quad
\sin2\theta_m=-{\sqrt{3}\Delta\over\sqrt{(\varepsilon_m-\lambda)^2+3\Delta^2}},\quad
\mP,\nonumber\\&&\cos2\tilde\theta_m=-~{\varepsilon_m-\lambda\over\sqrt{(\varepsilon_m-\lambda)^2+3\Delta^2}},\quad
\sin2\tilde\theta_m=-~{\sqrt{3}\Delta\over\sqrt{(\varepsilon_m-\lambda)^2+3\Delta^2}},\quad
\mm,\label{extremum}\nonumber\\ \end{eqnarray} where $\lambda$ is
the Lagrange multiplier which ensures the color neutrality
constraint (\ref{cnc}). The gap equation for color neutral
superconductivity reads
\begin{eqnarray}1={G_C\over
V}\left(\sum_\mm+\sum_\mP\right){1\over\sqrt{(\varepsilon_m-\lambda)^2+3\Delta^2}}.\label{gapeqn0}
\end{eqnarray} By setting $\lambda=0$, the gap equation for color
{\bf neutral} (not color {\bf symmetrical}) superconductivity is
obtained. Notice that we have been able to achieve color neutrality,
eq. (\ref{cneutral}), independently of the $\lambda$ value, without
introducing any extra Langrange multiplier, as opposed to what is
done in \cite{iida,buballa1}.  We stress, however, that the
implementation of (\ref{18}) is essential. Otherwise, the present
BCS vacuum can be reduced to the usual one, discussed in the next
section, by an appropriate color rotation.

At the extremum, the expectation value of $\hat K_{MFA}$ reduces to
\begin{eqnarray}\nonumber&\langle\hat
K_{MFA}\rangle&=\sum_{m\leq\Omega'}(2\varepsilon_m-E_m)+\sum_{m>\Omega'}
(\varepsilon_m-E_m)+V\,{3\Delta^2\over4G_C},\\&&
=3\sum_{m\leq\Omega'}\varepsilon_m+\sum_{\mm}(|\varepsilon_m|-E_m)+
\sum_{\mP}(\varepsilon_m-E_m)+V\,{3\Delta^2\over4G_C},
\end{eqnarray} where $
E_m=\sqrt{(\varepsilon_m-\lambda)^2+3\Delta^2}.$
In terms of proper canonical quasi particle operators $f_{im}$ such
that $\{f_{im},f_{jn}^\dagger\}=\delta_{ij}\delta_{mn}$, the
constrained Hamiltonian reads
\begin{eqnarray}&&\hat K_{MFA}=\sum_{m\leq\Omega'}\varepsilon_m+\sum_{m}
(\varepsilon_m-E_m)+V\,{3\Delta^2\over4G_C}+\sum_m\left(\sum_{j=1}^2
E_mf_{jm}^\dagger f_{jm}+ \varepsilon_mf_{3m}^\dagger
f_{3m}\right).\nonumber\\&&
\end{eqnarray}
As explained in
\cite{bohr}, the index $j$ in ${f_{jm}}$ does not specify a well
defined color. Indeed,
\begin{eqnarray*}f_{1m}=\kappa_1(d_{1m}-d_{2m}),\quad
f_{2m}=\kappa_2(d_{1m}+d_{2m}-2d_{3m}),\quad
f_{3m}=\kappa_3(d_{1m}+d_{2m}+d_{3m}),
\end{eqnarray*} where $\kappa_1,~\kappa_2,~\kappa_3,$ are
normalization constants.
\vskip.5cm \section{Comparison with usual color superconductivity}

In the usual approach to color superconductivity, which breaks color
symmetry \cite{alford,alford1}, we have
$\Delta_3=\Delta\neq0,~\Delta_1=\Delta_2=0$. The BCS transformation
which digonalizes $\hat K_{MFA}$ reads
$$c_{1m}=\alpha_m d_{1m}+ \beta_m d_{2\bar m}^\dagger,\quad c_{2\bar
m}=\alpha_m d_{2\bar m}- \beta_m d_{1 m}^\dagger, $$ with
$$
\alpha_m^2-\beta_m^2={\varepsilon_m\over\sqrt{\varepsilon^2_m+\Delta^2}},\quad
2\alpha_m\beta_m={\Delta\over\sqrt{\varepsilon^2_m+\Delta^2}},\quad
\alpha_m^2+\beta_m^2=1.
$$ The gap is expressed as $$\Delta=2{G_C\over
V}\sum_m\alpha_m\beta_m,
$$
so that, the gap equation reads
\begin{eqnarray}1={G_C\over
V}\sum_m{1\over\sqrt{\varepsilon^2_m+\Delta^2}}.\label{gapeqn}
\end{eqnarray}
The constrained mean field  Hamiltonian 
reduces to
\begin{eqnarray*}&\hat
K_{MFA}&=3\sum_{\mm}\varepsilon_m+\sum_{\mm}(|\varepsilon_m|-E_m)
+\sum_{\mP}(\varepsilon_m-E_m)+V\,{\Delta^2\over4G_C}\\&&+
\sum_{m}\left(E_m(d_{1m}^\dagger d_{1m}+d_{2 m}^\dagger d_{2
m})+\varepsilon_m c_{3m}^\dagger c_{3m}\right).
\end{eqnarray*}
The gap $\Delta$ is only due to the pairing correlations between
colors 1 and 2. The gap equation has essentially the same form in
the color symmetric (eq.(\ref{gapeqn0})) and in the color symmetry
breaking (eq. (\ref{gapeqn})) description of the color
superconductivity. However, in the first case we have automatically
$\langle S_{11}\rangle=\langle S_{22}\rangle=\langle S_{33}\rangle$,
while in the second case we have $\langle S_{11}\rangle=\langle
S_{22}\rangle\neq\langle S_{33}\rangle$. In \cite{buballa}, it has
been argued that the usual color superconductivity, where only two
colors participate in the gap, softens the equation of state. It is
expected that the participation of the three colors on the same
footing will have significant consequences for the high density
phases of QCD which exhibit color superconductivity. We do not
discuss the important color-flavor-locking mechanism, which also
leads to color neutrality, since our main concern is the two flavor
case.

\begin{figure}[ht]
\centering
\includegraphics[width=0.6\textwidth, height=0.5\textwidth]{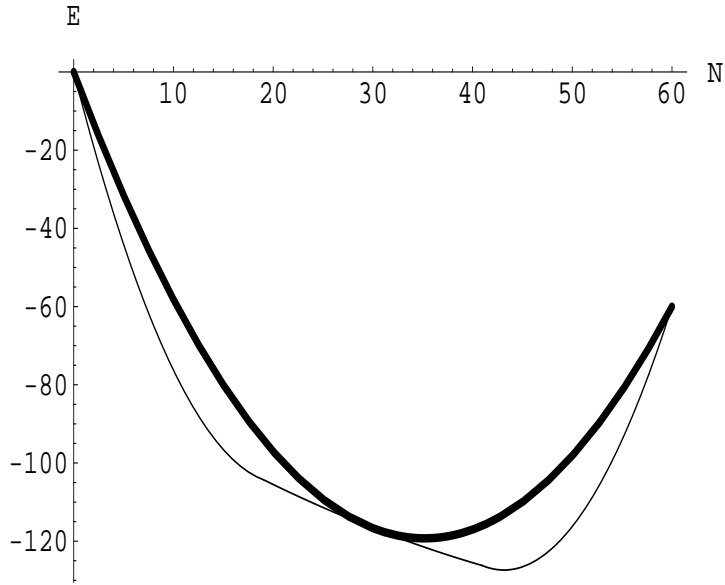}
\caption{Lowest energy in the BCS approximation for the schematic
Bonn model \cite{bohr} with $\Omega=10$, versus the quark number, in
arbitrary units. Thick line: the lowest energy of the color
symmetric sector, according to the new method presented in Section
2; thin line: the result of the usual approach, which is summarized
in Section 3, and describes the full groundstate energy of the
model, including the ``unphysical" sectors.} \label{fig1}
\end{figure}

\section{Conclusions}
Quarks in QCD are, in the confined phase, only allowed to form
colorless states. However, at high temperature and density, quarks
are expected to be free. It has been shown that in the deconfined
phase, which then prevails, the tendency for the formation of BCS
pairs occurs \cite{alford,alford1}. This gives rise to the so called
color superconducting phase, which, in the usual treatment
\cite{hatsuda,buballa,alford,alford1,iida,buballa1}, violates color
symmetry. In this paper, we show that color superconductivity is not
incompatible with color symmetry.  It will be interesting to
compare, in a realistic model of the NJL-type, the properties of the
phase described by the color symmetric version of color
superconductivity presently considered, with the corresponding
properties described by conventional color superconductiviy, which
drastically breaks color symmetry. This has been done for a
schematic model in ref. \cite{bohr}, and, there, the difference
found is important. In the Figure, we illustrate the performance of
the new method here propose, and compare it with the usual approach
to color superconductivity, in the context of the QCD inspired
schematic model considered in \cite{bohr}. This model, which is
characterized by a parameter $\Omega$ measuring the level
degeneracy, is admittedly unrealistic, but is quite useful to test
approximation techniques, since it is analytically solvable.
Applications to realistic situations are in progress.

 An effective QCD theory, as is appropriate to
describe e.g. the interior of neutron stars, will have vanishing
confining force at high temperatures and densities, due to
asymptotic freedom, but should also be consistent with color
singlet-ness \cite{gerhold,dietrich}. The phase of superconducting
color symmetric states is supposed to exist in the interior of
neutron stars with high density, where the simplified NJL model
becomes identical to an effective QCD field theory and, thus,
realistic.

The present approach is in contrast with those followed in refs.
\cite{amore} and \cite{iida}, where the color neutrality problem was
previously addressed. In \cite{amore}, the authors resort to rather
involved projection techniques to extract color neutral states out
of BCS sates which violate color symmetry. It should be pointed out
that the correlations described by the present approach need not
coincide with those arising within the framework of the projection
technique. In \cite{iida}, color neutrality is defined by the
condition that the average or expectation value of some of the eight
Gell-Mann operators vanishes, that is, color neutrality is
implemented with the help of appropriate Lagrange multipliers. In
connection with this question, a reference to a recent work of
Buballa and Shovkovy \cite{buballa1} is appropriate. These authors
observe that, if a common chemical potential $\mu$ is used for all
colors, then the quark numbers $\langle S_{11}\rangle,~\langle
S_{22}\rangle,~\langle S_{33}\rangle,$ are not all the same, and
hence will violate color neutrality. While, in the present approach,
equality of the quark numbers for different colors is achieved with
a common chemical potential, even if we set $\lambda=0$, in
\cite{buballa1} one advocates using different chemical potentials
for different colors, to insure $\langle S_{11}\rangle=\langle
S_{22}\rangle=\langle S_{33}\rangle$. However, in \cite{buballa1},
it is already observed that this condition is not sufficient to
insure color singlet-ness, due to its instability under color
rotations. We have proposed a solution to the emerging problem.

The present article also provides theoretical ``tools" for
constructing phases with superconducting color-symmetric states.

Although our aim is not to discuss the QCD Meissner effect, we
observe that this effect may be treated in terms of a Hamiltonian of
the form
$$H=\int\d^3x\left({1\over2}({\bf B}_a\cdot{\bf B}_a+{\bf E}_a
\cdot{\bf E}_a)+{\wt\Pi}^*\cdot{\wt\Pi}+ ({\bf
D}\wt\Delta)^*\cdot({\bf D}\wt\Delta)+V(\wt\Delta^*\cdot\wt\Delta)
\right),
$$
describing the color superconducting phase in interaction with the
gluon field. In the definition of the covariant derivative $\bf D$
one should keep in mind that $\wt \Delta$ belongs to the $\bar 3$
representation. It is clear that the standard treatment goes through
with our approach. We observe that our color symmetric BCS theory
should lead to qualitatively the same masses for the gauge bosons as
are found in the literature. The only place where an effective model
such as the NJL model comes in is in the estimation of the function
$V$ which depends on $|\Delta_1|^2,|\Delta_2|^2,|\Delta_3|^2.$ Our
approach ensures that $V$ will depend only on the combination
$|\Delta_1|^2+|\Delta_2|^2+|\Delta_3|^2.$

\section*{Appendix}
\subsection*{Derivation of eqs. (\ref{9}) and (\ref{10}).}
For $
\mP$, we have
$$X_m:=\langle c^\dagger_{1m}c_{2m}\rangle=-K^2_m-K^2_m\left(-\langle c^\dagger_{3\ol m}
c_{3\ol m}\rangle-\langle c^\dagger_{2\ol m} c_{1\ol m}\rangle
+\langle c^\dagger_{3\ol m} c_{1\ol m}\rangle+\langle
c^\dagger_{2\ol m} c_{3\ol m}\rangle\right)
$$
$$N_m:=\langle c^\dagger_{1m}c_{1m}\rangle=2K_m^2-K_m^2\left(\langle c^\dagger_{3\ol m}
c_{3\ol m}\rangle+\langle c^\dagger_{2\ol m} c_{2\ol m}\rangle
-\langle c^\dagger_{3\ol m} c_{2\ol m}\rangle-\langle
c^\dagger_{2\ol m} c_{3\ol m}\rangle\right),
$$ implying
\begin{eqnarray*}X_m=-K_m^2+K^2(N_m-X_m),\quad N_m=2K_m^2-2K_m^2(N_m-X_m),
\end{eqnarray*} which leads to $X_m=-K_m^2/(1+3
K_m^2),\;N_m=2K_m^2/(1+3 K_m^2). $ The corresponding expressions for
$ \mm$ are similarly obtained.
\subsection*{Derivation of eq. (\ref{14}).}
For $ \mP$ we have
$$\langle c_{2\ol m}c_{1
m}\rangle=K_m-K_m^2\left(\langle c_{2\ol m}^\dagger c_{3
m}^\dagger\rangle+\langle c_{3\ol m}^\dagger c_{1
m}^\dagger\rangle+\langle c_{1\ol m}^\dagger c_{2
m}^\dagger\rangle-\langle c_{3\ol m}^\dagger c_{3
m}^\dagger\rangle\right),
$$
$$\langle c_{1\ol m} c_{1 m}\rangle=-K_m^2\left(\langle c_{2\ol m}^\dagger c_{2 m}^\dagger \rangle
+\langle c_{3\ol m}^\dagger c_{3 m}^\dagger\rangle-\langle c_{3\ol
m}^\dagger c_{2 m}^\dagger\rangle-\langle c_{2 m}^\dagger c_{3\ol
m}^\dagger\rangle\right),
$$ which imply
$$D_m=K_m-3K_m^2D_m+K_m^2P_m,\quad P_m=2K_m^2P_m,
$$ where $P_m=\langle c_{1 m}c_{1\ol m}\rangle=\langle c_{2 m}c_{2\ol m}\rangle=\langle c_{3 m}c_{3\ol
m}\rangle,$ is also real. The procedure for $\Omega'<m$, is
analogous. Finally, we find $P_m=0 $ and {\color{red} eq. (\ref{14})
follows.}

 \subsection*{ Proof of eq.
(\ref{extremum}).} Let
$$\Psi(\theta,\tilde\theta,\Delta,\lambda)=\langle\hat
K_{MFA}\rangle+2\lambda\left(-\sum_{\mP}\sin^2\theta_m+\sum_{\mm}
\sin^2\tilde\theta_m\right),$$ where $\langle\hat K_{MFA}\rangle$
stands for its expression in eq. (\ref{17}) and $\lambda$ is a
Lagrange multiplier. We find
\begin{eqnarray*} &&\partial_{\theta_m}\Psi=4(\varepsilon_m-\lambda)\sin\theta_m\cos\theta_m
+2\sqrt{3}\Delta(\cos^2\theta_m-\sin^2\theta_m)=0,\quad\Omega'<m,\cr
&&\partial_{\tilde\theta_m}\Psi=-4(\varepsilon_m-\lambda)\sin\tilde\theta_m\cos\tilde\theta_m
+2\sqrt{3}\Delta(\cos^2\tilde\theta_m-\sin^2\tilde\theta_m)=0,\quad
m\leq\Omega'.
\end{eqnarray*}
The condition $\partial_{\Delta}\Psi=0 $ yields eq. (\ref{18}), and
the desired result, eq. (\ref{extremum}), follows now easily.


\section*{Acknowledgements}

The authors are grateful to the Referees for their constructive
criticism and valuable comments. The present research was partially
supported by Project 
CERN/FP/ 83505/2008.

\end{document}